%% file: main.tex
\documentclass[journal]{IEEEtran}

\usepackage[utf8]{inputenc}
\usepackage[T1]{fontenc}
\usepackage{authblk} 
\usepackage{graphicx}
\usepackage{amsmath, amssymb, amsfonts}
\usepackage{booktabs}
\usepackage{url}
\usepackage{hyperref}
\usepackage{xcolor}
\usepackage{geometry}
\usepackage{stfloats}
\usepackage{subcaption}

\usepackage[style=numeric, sorting=none, backend=biber]{biblatex}
\title{Securing the Sandbox: A Rootless Containerized Framework for Process-Oriented Monitoring in Computer Graphics Education}

\author[1]{Germán Arroyo}
\author[1]{Luis López}
\author[1]{Juan Carlos Torres}

\affil[1]{\scriptsize Departmento de Lenguajes y Sistemas Informáticos, University of Granada, Spain}

\date{\today}

\begin{document}

\maketitle

\begin{abstract}
Computer Science education fundamentally depends on intensive laboratory hours to foster true programming mastery and logical reasoning. However, the widespread adoption of Generative Artificial Intelligence (AI) has made it virtually impossible to distinguish authentic student effort from instant AI code synthesis by evaluating final submissions alone. To preserve pedagogical integrity, educators must enforce authentic coding discipline, guiding students through unassisted, iterative development cycles. While centralized environments like JupyterHub provide instructors with a platform to host and monitor the learning process step-by-step, they introduce severe operational vulnerabilities; because Jupyter environments inherently allow arbitrary shell command execution, they expose the underlying shared host to unauthorized system manipulation and lateral movement. This paper presents VISMATIC, a secure, low-cost framework designed to resolve this tension between process-oriented monitoring and infrastructure security. By pairing robust environment isolation with explicit user-interaction tracking at the API level, VISMATIC captures authentic programming behaviors without exposing the underlying host system. Evaluation from a pilot student cohort demonstrates that our macro-level behavioral metrics successfully flag statistical anomalies indicative of automated or off-platform workflows while preserving student anonymity, offering a scalable blueprint for safeguarding educational discipline in the AI era.
\end{abstract}

\input{sections/introduction}

\input{sections/previous_works}
\input{sections/methodology}
\input{sections/results}
\input{sections/conclusion}

\printbibliography

\end{document}

%% file: sections/introduction.tex
\section{Introduction}\label{sec:intro}

Computer Science education, particularly in specialized fields like Computer Graphics (CG), fundamentally relies on intensive laboratory hours to foster engineering discipline, mathematical reasoning, and true programming mastery. However, the rapid proliferation of Generative Artificial Intelligence (AI) tools has broken traditional grading paradigms; because sophisticated, syntactically correct graphics code can now be synthesized instantaneously, evaluating a student's final code submission alone no longer confirms authentic learning or individual effort. To safeguard pedagogical integrity, educators must shift their focus from the final product to the underlying development process, enforcing an unassisted, iterative coding discipline where students work through algorithmic hurdles step-by-step. Centralized development platforms, such as the JupyterHub ecosystem, offer professors a powerful method to orchestrate, log, and audit these learning workflows in real time. Yet, implementing this level of continuous oversight exposes institutional networks to a severe operational paradox: because interactive notebook environments inherently permit arbitrary shell command execution, opening a centralized server to remote student workloads provides a massive, multi-tenant attack surface vulnerable to unauthorized system manipulation, malicious internet downloads, and lateral network movement. In other words, the very infrastructure required to counter AI-assisted plagiarism---a centralized, monitored environment with remote execution capabilities---inherently introduces new security vulnerabilities that must themselves be addressed.

\begin{figure*}[htbp]
  \centering
  \includegraphics[width=\linewidth]{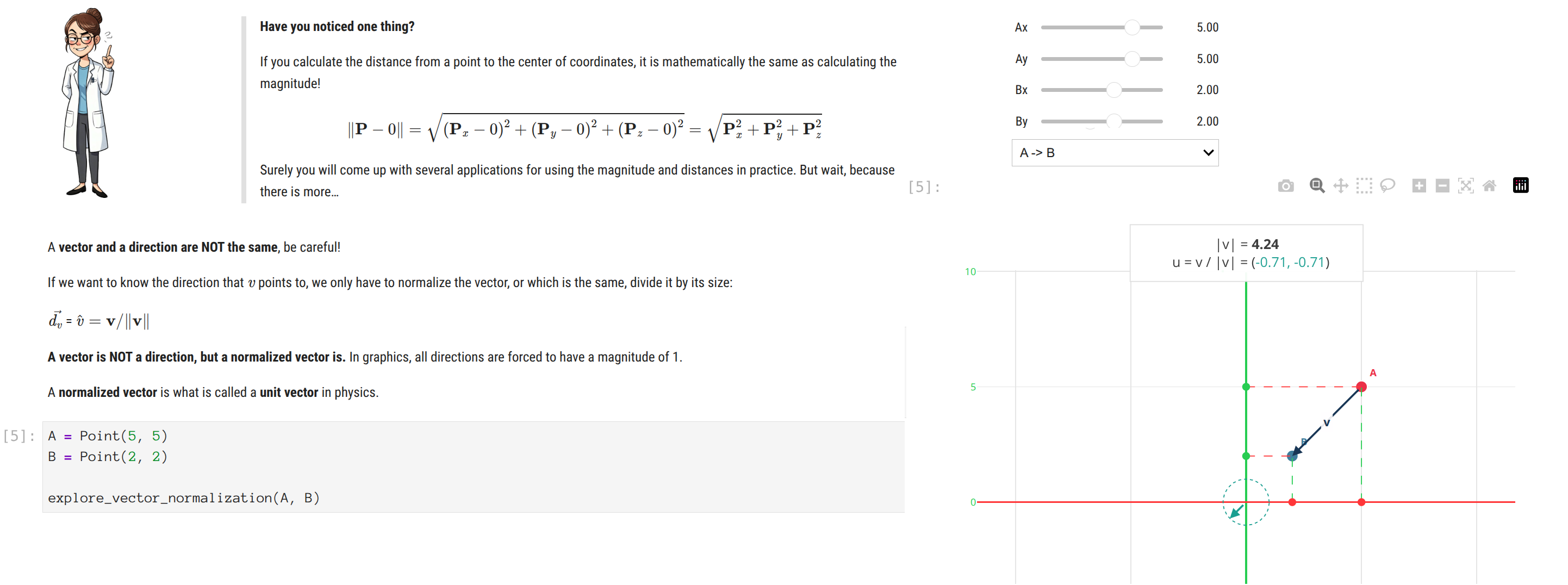}
  \caption{A typical student view of the VISMATIC Jupyter notebook interface. The environment seamlessly integrates \LaTeX-rendered mathematical theory with Python code and interactive visualizations, enabling students to manipulate variables and observe instantaneous graphical feedback within a secure, monitored sandbox.}
  \label{fig:screenshot}
\end{figure*}

To address this dual challenge---detecting AI-assisted work without exposing the institution to additional risk--- we introduce VISMATIC (Interactive Visualization of Mathematical Foundations for Computer Graphics). VISMATIC provides a hardened, "monitored sandbox" tailored for rigorous student inquiry and institutional containment. We emphasize that no system can achieve absolute security; our contribution is a defense-in-depth architecture that mitigates the specific risks introduced by deploying a monitored, multi-tenant execution environment. The foundational layer of the framework consists of interactive computational documents that integrate complex mathematical theory in \LaTeX\ with real-time graphical feedback. Within this ecosystem, students manipulate geometric transformations, lighting models, and rendering pipelines, observing changes instantaneously. Crucially, rather than relying on passive server heartbeats that risk inflating active student hours via background tabs, VISMATIC intercepts explicit HTTP API requests—specifically tracking file modifications, kernel executions, and terminal operations. This allows instructors to analyze macro-level interaction rhythms and immediately flag the sudden "copy-paste" injections characteristic of off-platform generative AI usage. The interface presented to students within the VISMATIC environment is illustrated in Figure~\ref{fig:screenshot}.

The core contribution of this work is a secure, layered framework that shifts the paradigm of Computer Graphics education from artifact-based grading to process-oriented supervision for undergraduate and master's courses. By chaining rootless containment with explicit API interception, the VISMATIC architecture safely hosts multi-tenant environments while building a transparent learning analytics pipeline. Our results validate that this framework effectively captures genuine student engagement dynamics and exposes automated anomalies—such as stay-alive scripts, macro usage, and off-platform AI code generation—that traditional submission-based grading fails to detect. Furthermore, we prove that by decoupling server-side orchestration from client-side graphics rendering, this framework can be deployed cost-effectively on resource-constrained hardware like the Raspberry Pi 5 without sacrificing operational stability.

This efficiency relies directly on our rootless container design. Unlike standard deployments that depend on host-level root privileges, our architecture guarantees that even if a student executes erratic shell commands—whether as an unvetted side effect of AI scripts or disjointed local workflows—the potential blast radius is strictly contained. Users lack the underlying operating system permissions to compromise the host kernel or cross into peer workspaces, ensuring total isolation and absolute traceability of macro-level actions at the point of origin. Ultimately, this project provides a scalable blueprint for modern STEM curricula where authentic student engagement can no longer be assumed, delivering an economical, tamper-proof telemetry infrastructure without burdening low-cost institutional hardware.

The remainder of this paper is organized as follows:

\begin{itemize}
\item Section~\ref{sec:previous} reviews the relevant literature regarding Generative AI challenges in computing education and the security concerns of multi-tenant server infrastructure.
\item Section~\ref{sec:methodology} delineates the technical implementation of our rootless isolation and container cluster deployment.
\item Section~\ref{sec:results} presents the behavioral metrics derived from our explicit API event tracking and evaluates the system's operational containment.
\item Section~\ref{sec:conclusion} summarizes our structural findings and details the pedagogical implications of secure process-oriented evaluation in AI-assisted environments.
\end{itemize}

%% file: sections/previous_works.tex
\section{Previous Works}\label{sec:previous}

While digital learning platforms initially evolved to resolve local deployment and environment fragmentation~\cite{Perez2007IPython, Kluyver2016Jupyter}, modern computer graphics education must instead navigate heavy inherent cognitive workloads~\cite{angel2015interactive, Chiodini2025Graphics} while countering the academic integrity risks posed by generative AI. Accordingly, this section reviews how remote interactive architectures leverage secure containerization and process analytics to safely enforce coding discipline without exposing shared host infrastructure to unauthorized command execution.

\subsection{Notebook-Based Learning}

The \textbf{Jupyter} ecosystem~\cite{Jupyter} has shifted the educational paradigm from static code submission to interactive, narrative-driven exploration~\cite{Kluyver2016Jupyter}. By utilizing domain-specific widgets~\cite{Du2024JupyterWidgets}, instructors can create environments that provide immediate visual feedback, bridging the gap between theory and practice. However, the value of these tools has shifted; while they were once valued for "lowering the barrier-to-entry"~\cite{apahidean2025containerized}, they are now essential for process-oriented evaluation. Frameworks like \textbf{nbgrader} automated the feedback loop~\cite{Hamrick2019nbgrader}, but modern requirements demand deeper telemetry. Recent advancements, such as \textbf{CodeDive}~\cite{Park2025CodeDive}, emphasize real-time activity monitoring as a means to ensure that the student—rather than an external AI—is the primary agent of code synthesis.

\subsection{Monitoring vs. Vulnerability}

The AI integrity crisis compels educators to monitor student work, yet the very act of monitoring creates a new vulnerability. The necessity of monitoring student progress in the AI era requires a transition to centralized, multi-tenant infrastructures. This shift introduces a critical security paradox: providing students with the remote execution capabilities required for auditing also grants them access to shared kernel resources, creating a significant attack surface. High-profile vulnerabilities such as "Copy Fail" (CVE-2026-31431) demonstrate that deterministic logic flaws in the kernel can bypass container boundaries entirely. By manipulating the shared page cache, an unprivileged user can trigger a host-wide privilege escalation or lateral movement into neighboring containers~\cite{Xint2026CopyFail}. This vulnerability is particularly acute in standard Docker environments, where the reliance on a root-privileged daemon and traditional UID mappings provides a broader path for an attacker to compromise the underlying host once an isolation boundary is breached.

To mitigate these risks, academic computing is increasingly adopting defense-in-depth strategies, specifically \textbf{rootless container orchestration}. By decoupling the container engine from the root user, institutions can provide the necessary "sandboxed" environment for student code without risking host-level compromise. Furthermore, this infrastructure allows for the integration of learning analytics—such as eBPF-based instrumentation or distributed activity tracking~\cite{Cai2025JupyterAnalytics, Park2025CodeDive}—to distinguish between iterative, human-led development and the "all-at-once" code injections characteristic of AI-assisted plagiarism~\cite{Kasneci2023ChatGPT, FinnieAnsley2022ChatGPT}.

\subsection{Gamification and Checkpoint-Based Integrity}

Gamification serves as both an engagement tool and a security framework. By decomposing complex graphics concepts into structured, game-like challenges~\cite{Deterding2011, jack2025exploring}, instructors can create "integrity checkpoints." For instance, tools like \textbf{QubitQuest} use mini-games to guide students through abstract logic~\cite{Hill2026QubitQuest}. In our framework, these gamified elements act as verifiable milestones; by requiring students to reach visual checkpoints through documented, iterative changes, we create a "proof-of-work" that is significantly harder to spoof with AI than a single final project submission~\cite{angel2015interactive}.

\subsection{Isolated Edge Computing for Risk Containment}

As established in the earlier architectural overview, the choice of hardware platform is integral to the security model. The hardware layer plays a vital role in institutional security. A \textbf{Raspberry Pi} cluster exemplifies this principle: it offers a method for physical and logical isolation at a low cost~\cite{Xu2023RaspberryPi}. By serving as the physical substrate for the rootless containers described above, it provides a self-contained, expendable compute node that can be dedicated exclusively to student workloads. By employing an "Edge" computing model~\cite{varghese2016challenges}, institutions can offload student workloads to dedicated, expendable nodes. This ensures that even in the event of a successful container breakout or resource-exhaustion attack, the impact is limited to a single low-cost node, protecting the broader university network while still providing students with sufficient power for rendering and numerical tasks. In this sense, the Raspberry Pi is not merely a cost-saving measure but a deliberate risk-containment strategy: it physically isolates the monitored environment, preventing lateral movement from a compromised student session to the institutional network.

%% file: sections/methodology.tex
\section{Methodology}\label{sec:methodology}

The VISMATIC framework is built upon a defense-in-depth architecture that mitigates the security risks introduced by process-oriented monitoring. By centralizing the execution environment on low-cost hardware, we provide the high-performance tools necessary for Computer Graphics (CG) education while containing each identified risk through a specific hardening strategy: rootless containerization to prevent privilege escalation, loop-device images backed by quotas to prevent resource exhaustion, and physically isolated hardware to prevent lateral movement.

\begin{figure*}[htbp]
  \centering
  \includegraphics[width=0.9\linewidth]{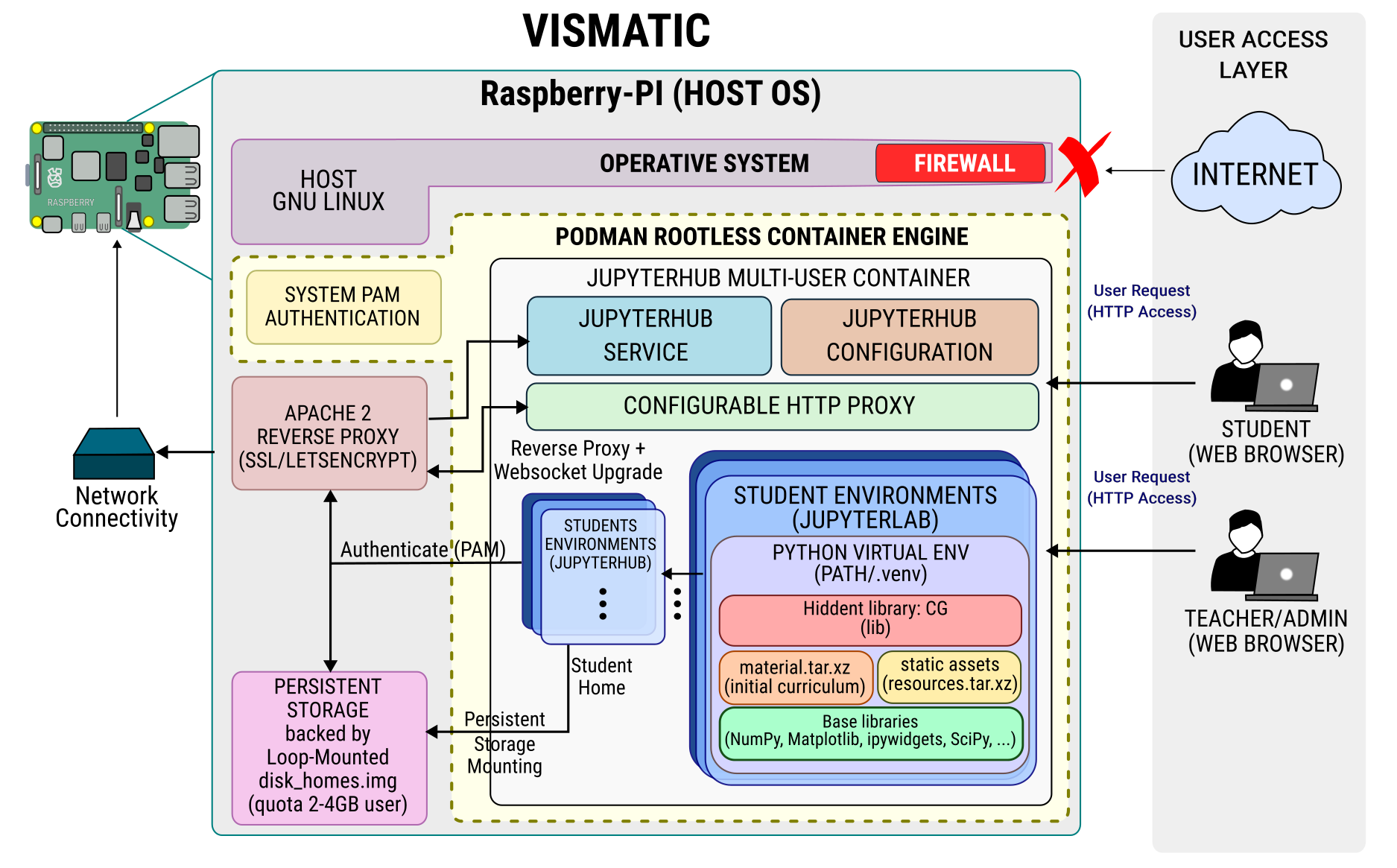}
  \caption{Architectural overview of the VISMATIC secured sandbox. The system employs a layered defense strategy: (1) an Apache2 reverse proxy for SSL termination and WebSocket tunneling, (2) a rootless Podman execution layer to isolate student processes from the host kernel, and (3) a loop-device filesystem backend to enforce physical storage quotas and ensure data persistence across container lifecycles.}
  \label{fig:diagram}
\end{figure*}

The structural containment of the VISMATIC framework is illustrated in Figure~\ref{fig:diagram}, which delineates the isolation boundaries between the public-facing proxy, the containerized execution environment, and the persistent storage abstraction.

\subsection{Infrastructure and Rootless Virtualization}

The most immediate risk introduced by granting students remote shell access is privilege escalation: a student might exploit kernel vulnerabilities or misconfigured permissions to escape the container and compromise the host. To mitigate this, deploying on a low-cost Raspberry Pi node limits physical infrastructure exposure but does not inherently address software-level privileges. To systematically prevent privilege escalation, the framework delegates all execution to a strictly non-privileged system user on the host. The primary line of defense relies on a rootless containerization layer paired with hardware-level namespace isolation.

This dual-layer protection of rootless containerization and namespace isolation is operationally realized through the use of Podman for container orchestration. Unlike traditional Docker-based solutions that rely on a root-privileged daemon, our implementation leverages Linux User Namespaces to map student processes to unprivileged UIDs on the host system. This design introduces a strong isolation boundary against container escape and kernel-level privilege escalation vulnerabilities, including recent Copy Fail (CVE-2026-31431)~\cite{Xint2026CopyFail}, Dirty Pipe~\cite{dirtypipe2026}, and Dirty Frag~\cite{v4bel_dirtyfrag_2026}. Even if a student manages to compromise the container environment, the resulting process remains confined to an unprivileged namespace with no administrative capabilities on the host, preventing modification of system binaries or interference with other users' environments, making the attacks much more difficult.

\subsection{Storage Abstraction and Quota Management}

A common vulnerability in multi-tenant educational environments is the "Disk Exhaustion" Denial-of-Service (DoS). To mitigate this, we implemented a loop-device storage abstraction. Student home directories are not hosted directly on the primary system partition; instead, they reside within a dedicated, fixed-size filesystem image formatted with \texttt{ext4}.

This approach provides two primary security benefits:
\begin{enumerate}
    \item \textbf{Deterministic Resource Limits:} By mounting the image via a loop device, we enforce an absolute physical ceiling on storage consumption. We further apply \texttt{setquota} utilities to manage individual student allocations, ensuring that no single user can impact the stability of the hardware, adding another security layer.
    \item \textbf{Persistence and Isolation:} Student data is decoupled from the ephemeral container lifecycle. While the computational environment can be destroyed and rebuilt instantaneously to maintain a "clean room" state, student work remains persistent within the isolated storage image.
\end{enumerate}

\subsection{Computational Stack and Integrity Checkpoints}

The execution environment is encapsulated in a custom Ubuntu 24.04-based container, optimized for the mathematical rigors of CG. To support process-oriented monitoring, the stack includes:
\begin{itemize}
\item \textbf{Centralized Gateway:} A JupyterHub instance manages multi-user sessions through a web-based interface, eliminating the risks and inconsistencies associated with local software installation and configuration on student devices.

\item \textbf{Domain-Specific Libraries:} A pre-configured Python environment provides scientific and interactive visualization tools, including \texttt{NumPy} for numerical computation and \texttt{ipympl}/\texttt{ipywidgets} for real-time interactive rendering within notebooks.

\item \textbf{Verification Utilities:} A read-only repository containing the official course materials is embedded into each container. A dedicated restoration utility (\texttt{restore}) enables students to synchronize their workspace with the reference version, simplifying maintenance and distribution of updated materials by instructors.

\item \textbf{CG Private Library:} A private custom library required to execute the practical exercises is deployed exclusively on the server side and remains inaccessible to students. As a result, students must remain connected to the platform to complete the activities, reducing the likelihood of offline completion or bulk AI-assisted code generation, while enabling teaching assistants and instructors to more effectively supervise student progress and interactions.
\end{itemize}

\subsection{Authentication and Network Security}

Centralizing student workloads on a networked server exposes the institution to network-level attacks, including unauthorized access to internal services and interception of student sessions. To mitigate these risks, student access is managed through the host system's Pluggable Authentication Modules (PAM) infrastructure, which is integrated into the Podman container environment. This allows for the creation of isolated system accounts in this paper that exist solely to facilitate the JupyterHub session. To secure the network perimeter, we employ an Apache2 reverse proxy with mandatory SSL encryption provided by Let's Encrypt.

The proxy configuration is specifically tuned for the WebSocket protocol, which is essential for the low-latency visual feedback required in CG education. This gateway ensures that all student-server interaction is encrypted and that the internal containerized infrastructure is never directly exposed to the public internet.

\subsection{Resource Enforcement and Telemetry}

To mitigate the risk of Denial-of-Service (DoS) through resource exhaustion—whether caused by malicious intent or poorly optimized student algorithms—the deployment framework enforces strict hardware orchestration via the container runtime. Each student session is restricted to a cap of 1.5 CPU cores and a maximum memory allocation of 0.5 GB (RAM). While these parameters are configurable based on the specific requirements of the rendering task, empirical testing demonstrated that this allocation provides a fluid user experience for Computer Graphics interactive notebooks. This resource efficiency is highly compatible with the pedagogical nature of core Computer Graphics curricula for undergraduate and master's students, which heavily emphasize foundational mathematics rather than brute-force algorithmic throughput. The primary learning curve for students centers on complex geometric and algebraic concepts—such as matrix multiplications, vector normalization, and polygon mesh orientation—where individual operations are computationally trivial. Even when advancing to rendering techniques like ray-casting, the primary difficulty lies in the mathematical derivation of ray-primitive intersections rather than high-compute scaling. Consequently, these essential concepts can be taught comprehensively on low-cost hardware without triggering performance bottlenecks. Under these constraints, a single Raspberry Pi node maintained high stability and responsive interactivity with a concurrent load of 10 to 20 students, ensuring the scalability of the low-cost infrastructure for a total annual throughput of over 300 students.

By centralizing execution within these resource-capped sandboxes, we enable the collection of interaction telemetry. These logs provide a high-fidelity view of the student's development path, allowing instructors to verify conceptual mastery and ensure that the "mathematical anxiety" is being solved through authentic inquiry rather than automated copy-pasting from AI assistants.

\subsection{Resource Optimization and Asset De-duplication}

Finally, to maintain high performance on the Raspberry Pi's microSD-based I/O, we implemented a shared asset strategy tailored to the specific constraints of multi-tenant Jupyter environments. In a standard JupyterHub deployment, user workspaces are strictly isolated within individual home directories; accessing files outside these boundaries often breaks the Jupyter UI file explorer or violates container security rules, which frequently forces instructors to naively duplicate large files across every student's storage. To circumvent this without compromising isolation, our framework implements read-only volume bind-mounts at the container runtime level. Large binary data—including high-resolution textures, instructional videos, and rendering datasets—are stored in a single, centralized directory mapped identically into each student's workspace by means of soft links. This allows the notebooks to safely utilize relative path referencing to a global, read-only repository. This de-duplication strategy significantly reduces the physical storage footprint and eliminates disk write contention during peak usage, ensuring that the limited I/O bandwidth of the single-board computer is strictly reserved for active student computation and telemetry logging.

%% file: sections/results.tex
\section{Results and Discussion}\label{sec:results}

For the experiments, students were informed that the material in the provided notebooks would form part of the final exam and was therefore considered examinable study content. The material was introduced briefly during lectures, with an estimated workload of approximately 4 hours per notebook to complete and master. Data was collected by intercepting explicit HTTP API requests within the JupyterHub environment. Initially, the platform's default telemetry heartbeat (POST /activity) was considered; however, because this mechanism acts as an automated browser keep-alive signal firing approximately every 300 seconds regardless of human interaction, it risks artificially inflating engagement times through idle background tabs. To ensure scientific rigor, automated telemetry was discarded. Instead, our methodology exclusively tracked explicit, human-driven API interactions: file modifications, kernel code executions, and terminal usage. While this approach cannot extract sub-second keystroke patterns, it provides a highly reliable measure of active macro-level study sessions and interaction rhythms without heavily burdening the I/O of resource-constrained hardware such as a Raspberry Pi 5. Crucially, the course utilizes a closed-source, environment-restricted Computer Graphics library. This architectural constraint prevents students from downloading the Jupyter notebooks for local, offline execution, thereby ensuring that all genuine iterative development, rendering, and debugging must occur persistently within the monitored platform.

The VISMATIC framework adopts a \textit{Privacy-by-Design} architecture that preserves student anonymity while maintaining the level of granularity required for behavioral monitoring and learning analytics. By decoupling the execution environment from institutional identity providers, the platform provides a secure and audit-ready infrastructure that protects student privacy from account provisioning through to final data analysis.

Unlike conventional educational platforms that rely on Single Sign-On (SSO) mechanisms or institutional identifiers, VISMATIC employs a proactive anonymization strategy. During infrastructure deployment, student accounts are provisioned through CSV-based initialization, enabling the automatic generation of randomized system credentials without requiring or storing personally identifiable information such as legal names or university identification numbers.

This pre-generated credential model ensures that all intercepted interaction data within the JupyterHub environment is inherently anonymized at the point of origin. For longitudinal dashboard analysis and reporting, the randomized accounts used in this study are mapped to persistent, animal-themed pseudonyms (e.g., \textit{Brave Dolphin}, \textit{Clever Kangaroo}). This dual-layer abstraction enables instructors and researchers to monitor the evolution of individual learning processes and coding rhythms over time, while maintaining a strict cryptographic separation between the behavioral event logs and official academic records, thereby strongly supporting institutional privacy requirements and data protection regulations.

\subsection{Interaction Data and Behavioral Metrics}

The VISMATIC framework was evaluated using a pilot cohort of 19 active students. Deployed on a resource-efficient Raspberry Pi 5 system, the platform recorded 1,880 discrete API events over approximately 57 active student-hours, resulting in a substantial dataset for behavioral, performance, and analytical research.

The monitoring capabilities of the framework derive directly from the centralized architecture of the JupyterHub environment. By enforcing the execution of all workloads on managed infrastructure rather than on local, unmonitored machines, the platform captures explicit interaction patterns that are typically inaccessible in traditional, decentralized development settings. As detailed in our methodology, the system avoids passive keep-alive telemetry and instead records three primary categories of explicit human-computer interaction:

\begin{itemize}
\item \textbf{File Operations (\textit{contents API}):} Events triggered when a student actively saves, modifies, or manages their Jupyter notebook files, indicating structural progress.
\item \textbf{Code Executions (\textit{sessions API}):} Events logged when a student sends code blocks to the kernel for processing, representing iterative testing, rendering, and debugging cycles.
\item \textbf{Terminal Usage (\textit{terminals API}):} Events capturing direct command-line interface interactions, typically utilized for environment management or advanced debugging.
\end{itemize}

To analyze the resulting logs, we defined four analytical metrics designed to differentiate authentic learning behaviors from automated or externally synthesized workflows.

\begin{figure}[hb]
  \centering
  \includegraphics[width=\linewidth]{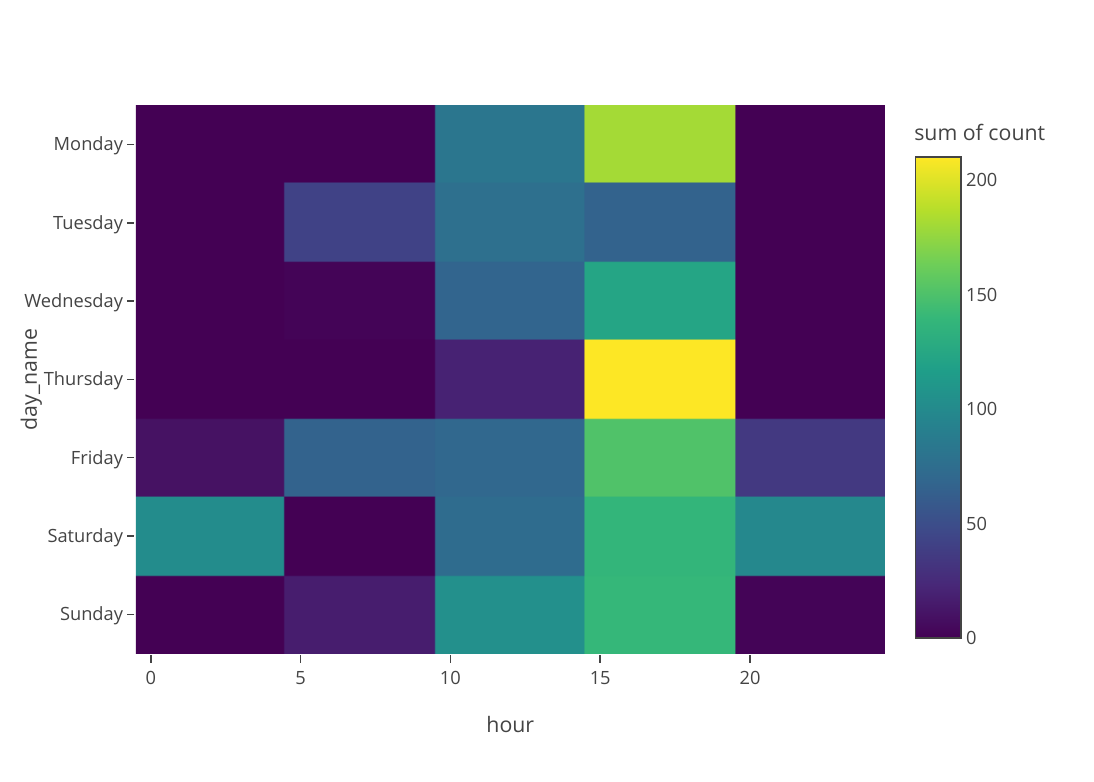}
  \caption{Daily activity heatmap illustrating temporal engagement patterns. The color gradient represents the frequency of explicit API interactions over time. The visualization contrasts distributed study behaviors with deadline-driven bursts of intensive activity.}
  \label{fig:heatmap}
\end{figure}

\paragraph{Class Rhythm Heatmap}
Represented in Figure~\ref{fig:heatmap}, this metric maps the temporal density of platform interactions across hours of the day and days of the week. This visualization reveals distinct diurnal and weekly work patterns within the student cohort. Peak activity consistently occurs during the evening hours (15:00 to 20:00), with Wednesdays exhibiting the highest concentration of interactions. Furthermore, the cohort demonstrates notable late-night to early-morning activity (00:00 to 05:00 and 20:00 to 00:00) specifically on Saturdays. These temporal signatures suggest a non-traditional student body that primarily engages with the platform outside of standard daytime hours—likely balancing academic coursework with professional commitments—and undertaking intensive, late-night programming sessions as the academic week progresses.

While the aggregate heatmap illustrates systemic cohort-level routines, applying this temporal metric on a per-student basis distinguishes between \textit{consistent learners}, who exhibit distributed activity patterns throughout the semester, and \textit{cramming behaviors}, characterized by concentrated late-night activity immediately before assignment deadlines. The educational relevance of this metric lies in its ability to associate temporal work habits with learning quality and conceptual assimilation.

High-density activity during late-night periods (e.g., between 00:00 and 04:00) is often associated with increased student stress and lower academic performance \cite{leinonen2021does}. In contrast, regular diurnal engagement patterns are generally indicative of a more structured and progressive approach to mastering the course material \cite{procaccino2002case}. Consequently, these temporal visualizations provide instructors with actionable insights into workload distribution, student well-being, and the pedagogical impact of assignment scheduling.

\begin{figure*}[p] 
\centering

    \begin{subfigure}{0.75\linewidth}
        \centering
        \includegraphics[width=\linewidth]{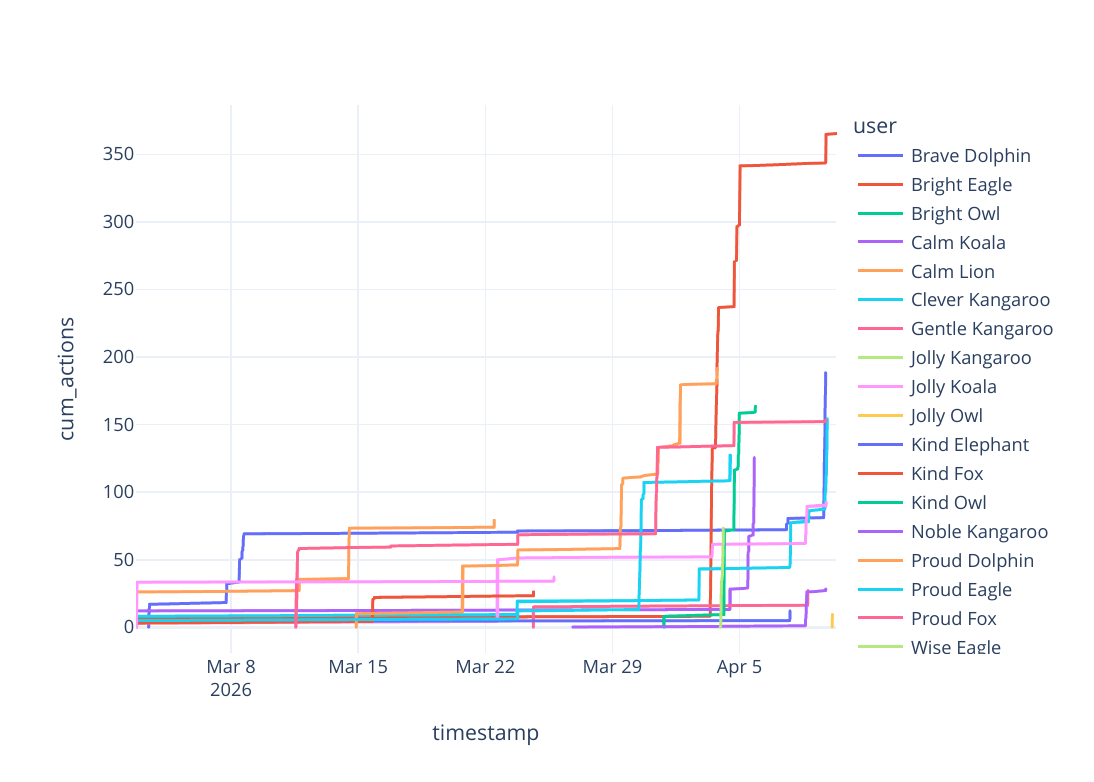}
        \caption{Analysis of cumulative platform engagement over time. The baseline illustrates the expected segmented growth associated with human iterative development, interspersing active coding with cognitive pauses. Anomalous unbroken, high-gradient linear trajectories indicate continuous automated interactions.}
        \label{fig:cumulative}
    \end{subfigure}

    \vspace{0.5cm} 

    \begin{subfigure}{0.75\linewidth}
        \centering
        \includegraphics[width=\linewidth]{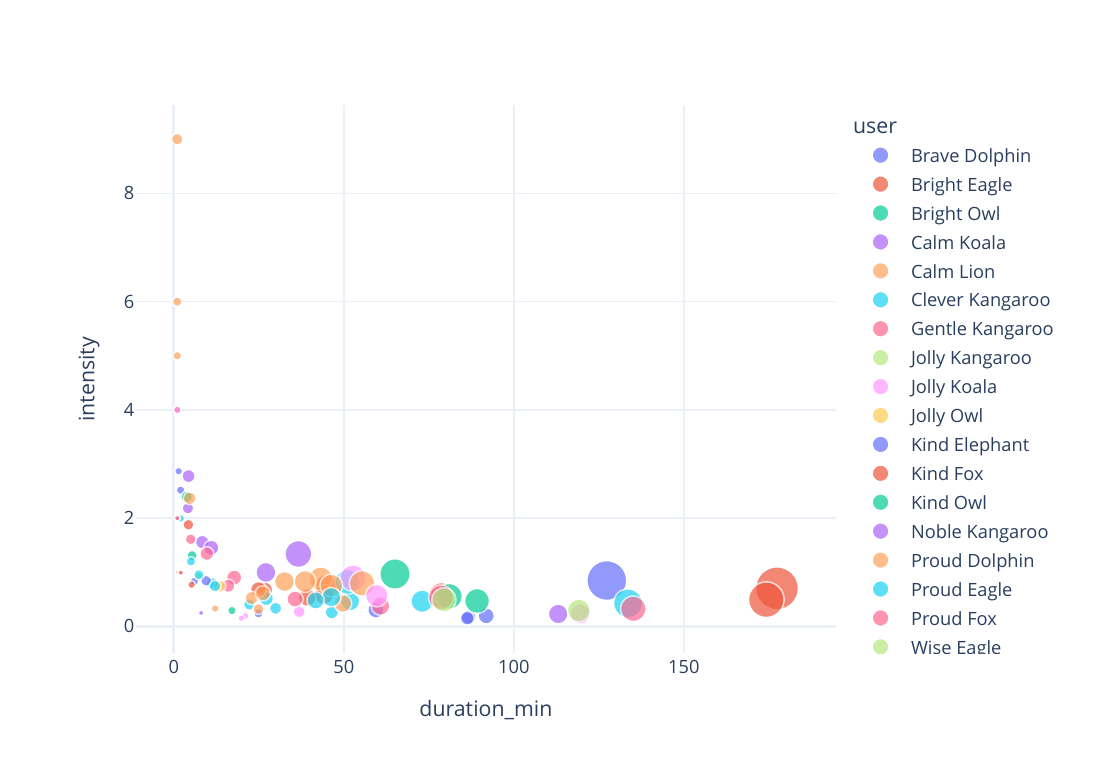}
        \caption{Distribution of session intensity versus duration across the student cohort. The left vertical cluster (high intensity, low duration) captures high-frequency burst activity, consistent with rapid code insertion or copy-paste operations from external workflows. Conversely, the far-right outliers characterized by extended duration and low event intensity indicate prolonged platform persistence, reflecting either distributed offline development cycles or automated attempts to maintain active container sessions.}
        \label{fig:intensity}
    \end{subfigure}

    \vspace{0.5cm}

\caption{Comprehensive behavioral analysis of platform interactions, highlighting human vs. automated development long term patterns.}
\label{fig:overall_analysis}
\end{figure*}

\paragraph{Cumulative Platform Engagement}
Figure~\ref{fig:cumulative} presents this metric, tracking longitudinal activity throughout the learning process. Because the underlying data extraction captures explicit API interactions (e.g., file modifications, code executions), this metric visualizes authentic, iterative development rather than passive server connection time. Genuine student engagement typically produces a segmented, staircase-like trajectory: active lab periods appear as phases of steady linear growth, which are naturally interspersed with extended flat plateaus. These plateaus correspond to offline periods, natural cognitive breaks, or time spent conceptualizing mathematical models outside the immediate platform environment.

The data exhibits clear ``staircase'' growth patterns, where cohort activity gradually scales upward as a deadline approaches. For example, platform engagement climbed steadily throughout early April, culminating in a massive peak of 348 aggregate events on April 4th, immediately followed by a sharp drop to baseline levels once the assignment was submitted. A secondary surge of 205 events occurred around April 9th prior to the final exercise deadline. This metric demonstrates that intercepted API event logs can accurately map the stress cycles and collective workload of a class, allowing instructors to identify periods of high academic pressure and potentially adjust future assignment distributions to encourage more consistent, rather than deadline-driven, learning habits.

\paragraph{Session Intensity}
This metric is defined as the ratio between the number of recorded explicit actions and the total duration of a reconstructed session ($Events/Time$). By relying on intercepted API requests rather than batched telemetry, this metric captures true physical engagement. A high intensity indicates rapid, iterative cycles of coding, rendering, and debugging, while a lower intensity suggests periods of reading, conceptualizing, or resting within a session. As shown in Figure~\ref{fig:intensity}, authentic human coding sessions naturally exhibit variable intensity.

Consequently, the framework uses this distribution to identify unnatural behavior: sessions that maintain an exceptionally high intensity with zero cognitive pauses over physically impossible durations are flagged as non-human automated behavior.

Accounts located in the extreme regions of the intensity-versus-duration distribution—such as continuous sessions spanning 24+ hours without a single 30-minute cognitive break to segment the session—are flagged as severe statistical outliers. Such behavioral profiles are physically impossible for human operators and are consistent with continuous server-side execution, browser macros, or automated ``stay-alive'' scripts hitting the platform APIs. By evaluating this macro-level persistence, instructors can clearly distinguish between organic, finite work cycles and unattended, synthetic interaction strategies.

\begin{figure*}[!h]
  \centering
  \includegraphics[width=0.75\linewidth]{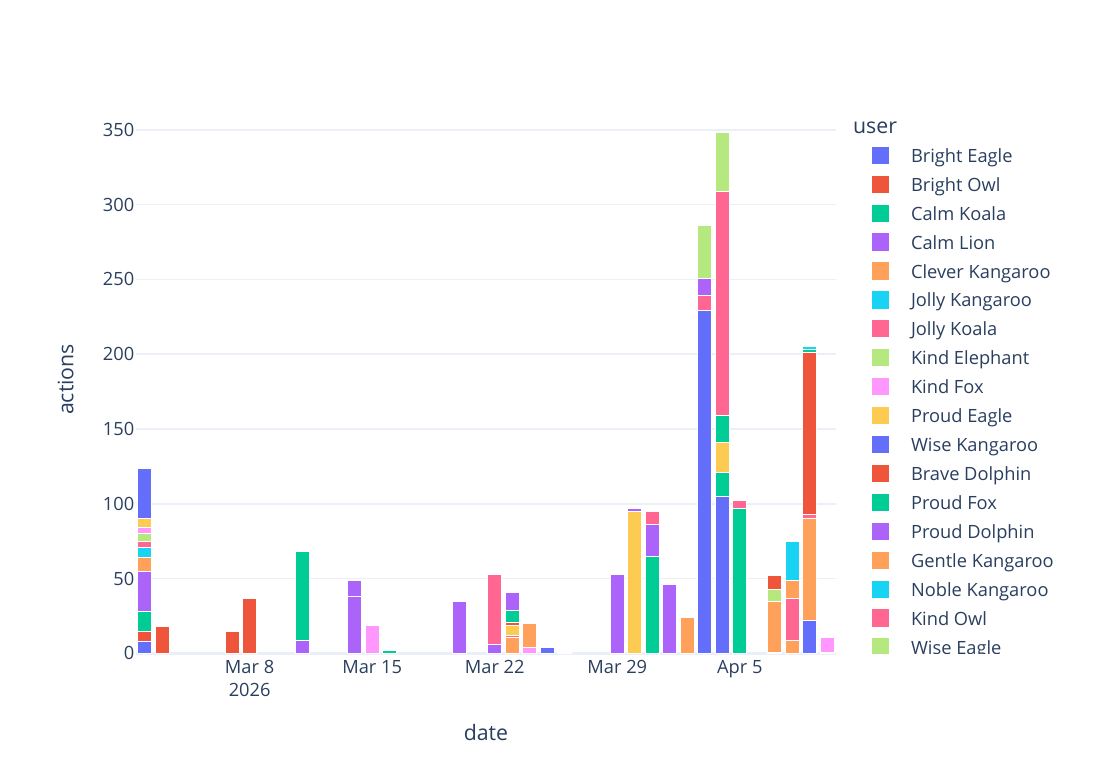}
  \caption{Longitudinal daily work volume. Aggregated explicit interaction events reveal the evolution of cohort engagement throughout the semester and highlight periods of collective activity.}
  \label{fig:volume}
\end{figure*}

\paragraph{Daily Work Volume}
This metric aggregates explicit interaction events over 24-hour periods to provide a longitudinal perspective of collective student engagement. As shown in Figure~\ref{fig:volume}, this macro-level analysis reveals the evolution of cohort engagement throughout the semester, beautifully visualizing the impact of assignment scheduling and deadline proximity. When triangulated with the cumulative engagement and session intensity metrics, daily volume provides the temporal context that anchors a comprehensive behavioral profile: while volume identifies \textit{when} the cohort is working under pressure, cumulative engagement clarifies \textit{how} this work is performed---iteratively with pauses versus linearly without rest---and session intensity reveals the \textit{nature} of the work within those spikes, distinguishing between sustained conceptual exploration and rapid, deadline-driven submission bursts.

\subsection{Anomaly Detection and AI Workflows}

By transitioning from passive server telemetry to explicit API interception, the proposed behavioral metrics successfully identified multiple anomalies consistent with automated workflows and off-platform code synthesis. Rather than relying on micro-level keystroke analysis, the framework successfully caught anomalous behavior through macro-level interaction deviations:

\begin{itemize}
    \item \textbf{Anomalous Session Persistence:} The framework detected unbroken sequences of API interactions spanning physically improbable durations. For instance, several statistical outliers maintained continuous code execution or file saving activity without a single 30-minute cognitive break for periods exceeding 24 hours. In these cases, the most plausible explanation is a browser macro or automated script interacting with the platform, as natural cognitive activity must include pauses and sleep cycles.

    \item \textbf{Absence of Temporal Plateaus:} Longitudinal profiles for flagged users exhibited nearly perfectly linear cumulative growth trajectories without horizontal resting periods. This rigid temporal signature is consistent with automated interaction patterns designed to simulate activity and prevent session termination.

    \item \textbf{Workflow Decoupling:} $Z$-score profiling was used to identify statistical outliers in the relationship between assignment completion and total active engagement time. A $Z$-score measures how many standard deviations an observation deviates from the mean of the dataset. Users with high completion rates but unusually low cumulative engagement time may indicate off-platform work (e.g., external tools such as generative AI or peer collaboration) followed by short, submission-oriented validation sessions within the monitored environment.
\end{itemize}

\subsection{Operational Security and Isolation}

\begin{figure*}[!htb]
  \centering
  \includegraphics[width=0.75\linewidth]{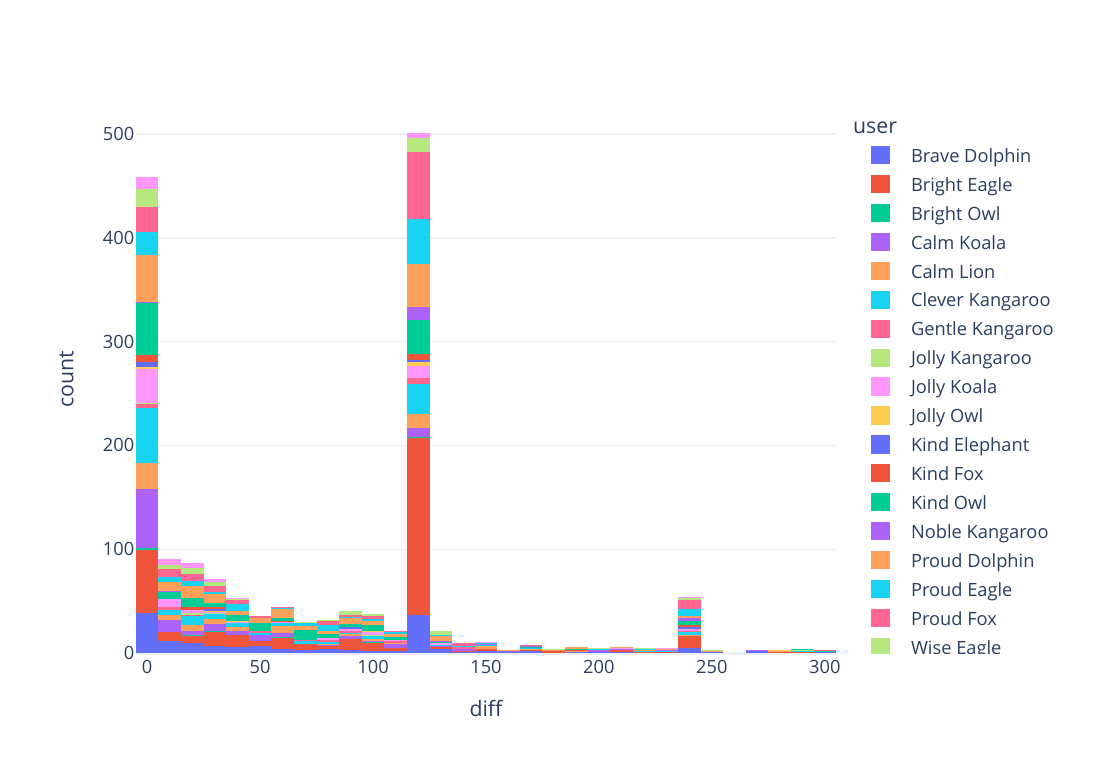}
  \caption{Distribution of time intervals between explicit keystrokes and click events. Short inter-event intervals correspond to high-frequency bursts of input activity, which may reflect rapid code entry, copy-paste behavior, or non-interactive code insertion patterns.}
  \label{fig:activity_rhythm}
\end{figure*}

To evaluate the operational security and isolation properties of the VISMATIC deployment, we analyzed system-wide API event logs under a single-tenant configuration. The dataset captures interactive computation and system-level session lifecycle events, including kernel initialization, code execution, and file operations. Across the observation window, interaction logs exhibit four dominant regimes of activity magnitude:

To further characterize student interaction patterns beyond aggregate activity counts, we analyze the time intervals between explicit keystrokes and click events (as shown in Figure~\ref{fig:activity_rhythm}).

This metric captures the fine-grained rhythm of user interaction and is particularly useful for identifying burst-like behavior, including rapid input sequences consistent with copy-paste operations or externally generated code insertion. Unlike cumulative event counts, inter-event timing provides a higher-resolution view of engagement dynamics, enabling a clearer distinction between sustained problem-solving activity and short, discontinuous input bursts.

\paragraph{Low-frequency initialization clusters (1--5 actions)}{
These values appear consistently across multiple users on specific dates. This regime is consistent with short-lived sessions or partially completed interaction windows, where computational environments are provisioned but not sustained. The homogeneity of values across users suggests standard environment bootstrap behavior rather than meaningful divergence in individual learning activity.
}

\paragraph{Standard operational workload cluster (6--44 actions)}{
The majority of active study blocks between March and May fall within this range. This stability indicates sustained JupyterHub sessions with continuous kernel interaction and typical development workflows. This regime dominates the dataset, suggesting the system supported steady, concurrent computational workloads from multiple users over extended periods without observable degradation or I/O saturation effects.
}

\paragraph{High-frequency iterative bursts (45--125 actions)}{
These values occur sporadically and are consistent with intensive development cycles, including rapid debugging, iterative testing, and frequent kernel restarts. Their presence is expected in interactive programming workflows, particularly in graphics- or computation-heavy exercises where short feedback loops are common.
}

No evidence of cross-identity contamination or simultaneous multi-session identity collision was observed in the dataset. Each user label remains temporally consistent within expected session boundaries, and no anomalous synchronization patterns (e.g., coordinated execution bursts across identities) were detected.

From a security perspective, the observed variability is consistent with the expected lifecycle of containerized educational workloads rather than external compromise. Although the system was exposed to continuous network access over an extended period, no indicators of privilege escalation, inter-user interference, or boundary violations were identified. Overall, the results suggest that the rootless Podman isolation layer maintained effective containment of execution environments, with workload variation primarily driven by normal differences in user interaction patterns.

\subsection{Discussion}

The results obtained from the VISMATIC deployment demonstrate that process-oriented monitoring can be integrated into Computer Graphics education without sacrificing either infrastructure security or student privacy. More importantly, the intercepted API data suggests that behavioral analysis provides a substantially richer representation of student learning than traditional submission-based evaluation methods.

The observed interaction patterns reinforce the hypothesis that authentic programming behavior exhibits identifiable temporal characteristics. Iterative workflows were consistently associated with gradual cumulative progress, variable session intensities, and distinct diurnal sleep cycles. In contrast, unnaturally rigid interaction rhythms, unyielding cumulative growth, and physically impossible persistence frequently corresponded to behaviors indicative of automated scripts or off-platform development. While these metrics cannot independently prove academic misconduct, they provide instructors with a probabilistic framework for identifying anomalous workflows that warrant further pedagogical review. The inherent limitation is that API-level event logs capture interaction patterns but not intent; a student may leave an automated script running for practical reasons unrelated to dishonesty, or may be genuinely working offline and later uploading solutions. To move from anomaly detection to corroborated evidence, instructors should triangulate these automated flags with additional assessment layers. For example, follow-up oral interviews with flagged students can probe conceptual understanding of the submitted code; comparing exam performance against platform activity can reveal mismatches between demonstrated ability and logged effort; and targeted plagiarism detection on notebook cells can identify code blocks that match known generative AI outputs. These complementary approaches preserve the value of the telemetry as a screening tool while acknowledging that no single metric can substitute for direct evaluation of student competence.

From an educational perspective, the framework shifts the focus of assessment from the final artifact toward the development process itself. This distinction is increasingly relevant in the context of Generative AI, where syntactically correct solutions can be produced instantly without meaningful conceptual understanding. By monitoring session persistence, temporal engagement rhythms, and workload distribution, VISMATIC captures aspects of student effort that are invisible in conventional grading pipelines. This process-oriented perspective aligns particularly well with STEM education, where learning often emerges through experimentation, visualization, and repeated mathematical adjustment.

The infrastructure results are equally significant. The rootless Podman architecture successfully isolated user workloads under realistic classroom conditions while maintaining stable performance on low-cost Raspberry Pi hardware. The absence of host-level compromise, lateral movement, or severe resource exhaustion events—even under the strain of high-frequency API logging—suggests that security-oriented educational infrastructures can be deployed economically without relying on enterprise-scale hardware. Furthermore, the use of loop-device storage abstraction and deterministic resource quotas proved effective in preventing common denial-of-service scenarios associated with shared educational servers.

Our results validate that by accurately measuring student dedication over time, this framework effectively deters students from taking shortcuts, such as relying on off-platform AI code generation or automated stay-alive scripts. By exposing these behavioral anomalies, the architecture provides professors with objective proof of genuine engagement, significantly mitigating the need for time-consuming individual interviews to verify lab work authenticity. Furthermore, we demonstrate that by focusing assignments on the mathematical foundations of graphics, this secure framework can be deployed cost-effectively on single-board computers like the Raspberry Pi 5, while remaining entirely modular and scalable to higher-performance cloud nodes as curriculum demands evolve.

%% file: sections/conclusion.tex
\section{Conclusion}\label{sec:conclusion}

This paper presented VISMATIC, a defense-in-depth framework that demonstrates how the security risks inherent to process-oriented monitoring can be effectively managed in Computer Graphics education. The proposed architecture combines rootless containerization, centralized JupyterHub orchestration, and low-cost Raspberry Pi infrastructure to mitigate the specific vulnerabilities---privilege escalation, resource exhaustion, and lateral movement---that arise when students are granted remote execution capabilities for integrity-preserving assessment.

Our results demonstrate that behavioral telemetry collected from centralized notebook environments can provide meaningful insight into the learning process beyond conventional submission-based evaluation. Metrics such as temporal velocity, inter-activity intervals, session intensity, and work composition enabled the identification of interaction patterns associated with iterative human problem solving as well as anomalous workflows consistent with external AI-assisted synthesis. Rather than focusing exclusively on final outputs, the framework emphasizes the importance of observing the incremental development process, which is particularly relevant in Computer Graphics education where experimentation, debugging, and mathematical refinement are central to conceptual understanding.

From a systems perspective, the deployment validated the feasibility of using rootless Podman containers as a defense-in-depth mechanism for educational infrastructures. The isolation guarantees provided by user namespaces, combined with deterministic resource limits and loop-device storage abstraction, successfully mitigated risks related to privilege escalation, lateral movement, and resource exhaustion. Despite the constrained hardware profile of the Raspberry Pi cluster, the platform maintained stable and responsive operation under realistic classroom workloads, demonstrating that secure remote laboratory environments can be implemented at low cost and with high energy efficiency.

The proposed framework also contributes to current discussions on the role of AI in STEM education. Instead of attempting to prohibit generative tools entirely, VISMATIC provides instructors with mechanisms to contextualize and interpret student behavior through longitudinal telemetry. This enables a transition from purely artifact-based assessment toward process-aware pedagogical evaluation. We do not claim absolute security; rather, we present a scalable risk-management approach that significantly hardens the infrastructure against the specific threats introduced by monitored remote learning, while acknowledging that security is an ongoing process requiring continuous adaptation.

Future work will focus on expanding the telemetry pipeline through machine learning and sequence-analysis techniques capable of modeling learning trajectories in greater detail. Additional research will also investigate adaptive interventions, automated anomaly classification, and the integration of privacy-preserving analytics to further strengthen the balance between educational visibility, student autonomy, and infrastructure security.